# Precise determination of the isotope shift of $^{88}$Sr – $^{87}$Sr optical lattice clock by sharing perturbations

Tetsushi Takano[1], Ray Mizushima[1], and Hidetoshi Katori[1,2,3*]

[1]*Department of Applied Physics, Graduate School of Engineering, The University of Tokyo, Bunkyo-ku, Tokyo 113-8656, Japan*

[2]*Quantum Metrology Laboratory, RIKEN, Wako-shi, Saitama 351-0198, Japan*

[3]*RIKEN Center for Advanced Photonics, Wako-shi, Saitama 351-0198, Japan*

E-mail: katori@amo.t.u-tokyo.ac.jp

We report on the isotope shift between $^{88}$Sr and $^{87}$Sr on the $^1S_0 - {}^3P_0$ clock transitions. The interleaved operation of an optical lattice clock with two isotopes allows the canceling out of common perturbations, such as the quadratic Zeeman shift, the clock-light shift, and the blackbody radiation shift. The isotope shift is determined to be 62 188 134.004(10) Hz, where the major uncertainty is introduced by the collisional shift that is distinct for each isotope. Our result allows us to determine the frequency of $^{88}$Sr-$^{87}$Sr optical lattice clocks with a fractional uncertainty of $2\times10^{-17}$. The scheme is generally applicable for measuring the isotope shift with significantly reduced uncertainty.

Optical clocks[1-7] outperform state-of-the-art Cs clocks by two orders of magnitude and are pressing the redefinition[8] of the second. Therefore, it has become essential to determine clock frequencies in terms of the frequency ratio $R$ such as $\nu(^{171}\text{Yb})/\nu(^{87}\text{Sr})$[9] and $\nu(^{199}\text{Hg})/\nu(^{87}\text{Sr})$[10, 11], allowing the transition frequencies to be described beyond the uncertainty of the SI second. Optical lattice clocks using $^{87}\text{Sr}$ play a pivotal role in these comparisons, as their frequency $\nu(^{87}\text{Sr}) = 429\ 228\ 004\ 229\ 873.2(2)$ Hz (Ref. 12) has been determined to an uncertainty of $5\times10^{-16}$, given solely by the realization of the definition of the SI second itself.

Optical lattice clocks with $^{88}\text{Sr}$ atoms[13-16], on the other hand, are widely developed because of their simple implementation, which is particularly suitable for building transportable clocks[17]. However, the absolute frequency of the clock transition in this isotope has only been determined with an uncertainty of $5\times10^{-15}$ (Ref.15), mainly limited by the quadratic Zeeman shift, clock-light shift, and lattice-light shift. In this Letter, we demonstrate the rejection of such perturbations by the interleaved interrogation of two isotopes and report on the most precise isotope shift $\Delta\nu_0 = \nu(^{88}\text{Sr}) - \nu(^{87}\text{Sr})$ that allows the determination of the $^{88}\text{Sr}$ clock transition frequency with a fractional uncertainty of $2\times10^{-17}$ with respect to $^{87}\text{Sr}$ clocks.

While the lack of nuclear spin ($I = 0$) in $^{88}\text{Sr}$ makes the clock insensitive to the lattice-light polarization effects[14], it requires a magnetic mixing field $\mathbf{B}_\text{m}$ to allow the $^1S_0 - {}^3P_0$ clock transition to occur with the Rabi frequency $\Omega_\text{R} = \alpha_\text{m}\sqrt{I_\text{R}}(\mathbf{B}_\text{m} \cdot \mathbf{e})$, where $I_\text{R}$ is the clock laser intensity with the polarization vector $\mathbf{e}$. The small coupling coefficient[18] $\alpha_\text{m} = 198\ \text{Hz}/(\text{T}\sqrt{\text{mW/cm}^2})$ necessitates a large mixing field $\mathbf{B}_\text{m}$ or clock laser intensity $I_\text{R}$ and introduces a relatively large quadratic Zeeman shift and clock-light shift, which constitute major systematic uncertainties for the clock.

Despite these difficulties, a three-dimensional (3D) optical lattice clock[14] with $^{88}$Sr is a promising candidate for realizing high stability and accuracy, as a singly occupied 3D lattice allows an increasing number of atoms while suppressing collisions. Hyper-Ramsey spectroscopy[6, 19] will be implemented to reduce clock-light shift uncertainty.

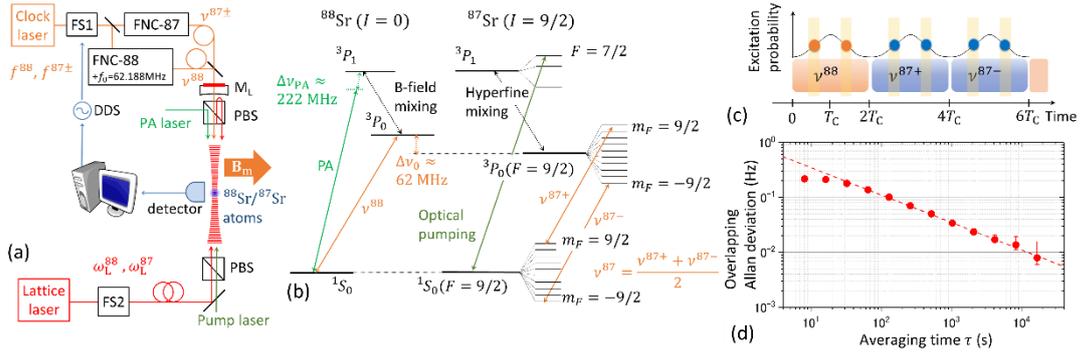

**Fig. 1.** (a) Experimental setup. $^{88}$Sr and $^{87}$Sr atoms are alternately loaded into an optical lattice. To reduce the collisional shift, we apply a photoassociation (PA) laser for $^{88}$Sr and a pump laser to spin-polarize $^{87}$Sr. The same magnetic mixing field $\mathbf{B}_\mathrm{m}$ and the off-resonant clock laser $\nu^{88}$ are applied to $^{87}$Sr to share perturbations between the two isotopes. The frequency shifter (FS1) steers the clock laser frequency using a computer-controlled direct digital synthesizer (DDS). A clock laser is sent to atoms via two phase-stabilized fibers. A lattice mirror (M$_\mathrm{L}$), which forms the lattice standing wave, defines the reference surface (indicated by a red line) for the fiber noise cancellers (FNCs). The two FNCs generate an offset frequency $f_0 = 62.188$ MHz. Lattice frequencies $\omega_\mathrm{L}^{88(87)}$ for the two isotopes are set by FS2. (b) Energy diagrams for $^{88}$Sr and $^{87}$Sr with clock transition frequencies ($\approx 429$ THz) of $\nu^{88}$, $\nu^{87} = (\nu^{87+} + \nu^{87-})/2$, and the isotope shift $\Delta\nu_0 = \nu^{88} - \nu^{87}$. (c) Time chart for the measurement. We repeat the measurements with the cycle time $T_\mathrm{C}$

to determine the excitation probability of the clock transition. Both shoulders of the Rabi excitation profiles for $^{88}$Sr and $^{87}$Sr ($m_F = \pm 9/2$) are probed in $6T_C$, and the correction frequencies $f^{88}$ and $f^{87\pm}$ are generated to drive FS1. (d) Overlapping Allan deviation for the isotope shift measured with $T_C = 1.35$ s and $\tau_R = 0.4$ s

Our experimental setup and the energy diagram for relevant transitions are shown in Figs. 1(a) and 1(b), respectively. $^{88}$Sr atoms are laser cooled using a two-stage magneto-optical trap (MOT)[20] and loaded into a 1D optical lattice consisting of a standing wave of a lattice laser tuned to $\omega_L^{88}$. We then apply a photoassociation (PA) laser, which is red-detuned by $\Delta\nu_{PA} \approx 222$ MHz[21] from the $^1S_0 - {}^3P_1$ resonance with an intensity of about $100$ W/cm$^2$, for a duration of 140 ms, to reduce the number of multiply populated lattice sites. On the other hand, $^{87}$Sr atoms with nuclear spin $I = 9/2$ are trapped in the lattice tuned to $\omega_L^{87}$ as described previously[4]. We then pump the atoms to the $m_F = \pm 9/2$ magnetic sublevels using a $\pi$-polarized laser resonant to the $^1S_0(F = 9/2) - {}^3P_1$ ($F = 7/2$) transition. Finally, we excite one of the Zeeman components to the $^3P_0(m_F = \pm 9/2)$ state and blow away atoms remaining in the $^1S_0$ state. Typically, we trap a few hundred $^{88}$Sr or spin-polarized $^{87}$Sr atoms distributed over $10^3$ sites in the 1D optical lattice for an average occupancy of $\bar{n} < 1$.

The isotope shift is measured by an interleaved clock operation with $^{88}$Sr and $^{87}$Sr atoms. A clock laser[22] at 698 nm is sent to the atoms with the polarization vector $\mathbf{e}(||\mathbf{B}_m)$ via two fibers that are phase-stabilized to a mirror [M$_L$ in Fig. 1(a)] forming the standing wave for the optical lattice with the polarization vector $\mathbf{e}_L(||\mathbf{B}_m)$. The fiber-noise cancellation systems FNC-88 and FNC-87 are configured such that they yield a frequency offset $f_0 = 62.188$ MHz that largely compensates the isotope shift $\Delta\nu_0$. In the experiment, all relevant RF sources are referenced to a calibrated hydrogen maser. A Rabi pulse with a duration of $\tau_R = 100 - 400$ ms sequentially interrogates [see Fig. 1(c)] the $^1S_0 - {}^3P_0$ transition of $^{88}$Sr and the $^1S_0(m_F = \pm 9/2) - {}^3P_0\ (m_F = \pm 9/2)$ transitions of $^{87}$Sr, corresponding to the frequencies $\nu^{88}$ and $\nu^{87\pm}$ [see Fig. 1(b)], with a cycle time of $T_C = 1.35 - 1.55$ s. The excitation probabilities are used to steer the clock laser frequencies by applying the correction frequencies $f^{88}$ and $f^{87\pm}$ to FS1. The isotope shift is measured as $\Delta\nu = f_0 + f^{88} - (f^{87+} + f^{87-})/2$. The overlapping Allan deviation for $\Delta\nu$ with $T_C = 1.35$ s and $\tau_R = 400$ ms is shown in Fig. 1(d), and improves as $\sigma_\nu(\tau) = 1.2$ Hz/$\sqrt{\tau/s}$ with the average time $\tau$.

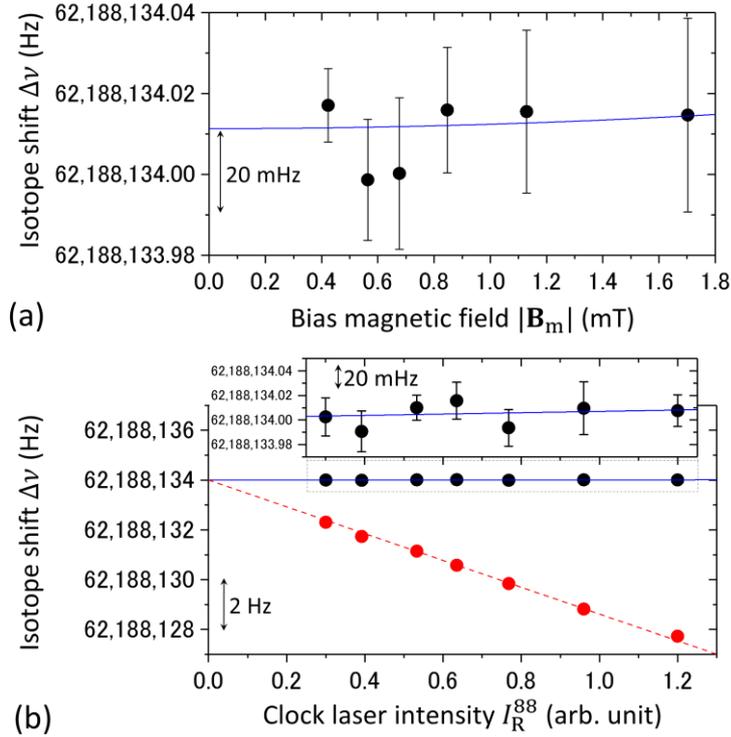

**Fig. 2.** (a) Isotope shift as a function of applied mixing field $|\mathbf{B}_m|$. The horizontal axis is measured by the first-order Zeeman shift $\nu^{87+} - \nu^{87-}$ between the magnetic sublevels $m_F = \pm 9/2$ of $^{87}$Sr. The two isotopes share

the same magnetic field to cancel out the quadratic Zeeman shift in deriving the isotope shift $\Delta v$ (black circles). The blue line shows a parabolic fit to investigate the residual Zeeman shift. (b) Isotope shift dependent on the clock laser intensity $I_R^{88}$, which causes significant clock-light shift (red circles). By applying the same clock laser (detuned by $f_0 = 62.188$ MHz from $v^{87}$) to $^{87}$Sr, we cancel out the clock shift, as shown by black circles. The error bars represent the $1\sigma$ statistical uncertainty obtained from the overlapping Allan deviation.

We apply the same magnetic field $\mathbf{B}_m$ for $^{88}$Sr and $^{87}$Sr to share the quadratic Zeeman shift $\delta v_Z^A = \beta^A |\mathbf{B}_m|^2$, which is expected to be a common perturbation for both isotopes with $\beta^A \approx -23.8(3)$ MHz/T$^2$ (Ref. 5). Hereafter, the superscript $A = 87$ and 88 denotes the mass number of the isotopes. The black circles in Fig. 2(a) show the isotope shift $\Delta v = \Delta v_0 + \beta^{88}|\mathbf{B}_m|^2 - \beta^{87}|\mathbf{B}_m|^2$ as a function of the mixing field $|\mathbf{B}_m|$, where we fix the clock laser intensity $I_R^{88} \approx 360$ mW/cm$^2$ and vary the duration of the Rabi pulse as $\tau_R = (2\Omega_R)^{-1} \propto |\mathbf{B}_m|^{-1}$. Note that a typical quadratic Zeeman shift is as large as $\delta v_Z^A \approx -17$ Hz for $|\mathbf{B}_m| = 0.85$ mT. By fitting a parabola (blue line), the fractional difference in the quadratic Zeeman shift coefficient is measured to be $\Delta \beta / |\beta| = (\beta^{88} - \beta^{87})/|\beta^{87}| = 0.5(2.0) \times 10^{-4}$, which is consistent to zero within the uncertainty.

Similarly, we investigate the clock-light shift $\delta v_R^A = \kappa^A(v) I_R^A$ induced by the clock laser with the intensity $I_R^{88}$ and the frequency $v$ for the differential dipole polarizability $\kappa^A(v) (\approx -18 \, \text{mHz}/(\text{mW/cm}^2))$[18]. Here, we set $\mathbf{B}_m = 0.85$ mT and vary the Rabi pulse duration as $\tau_R \propto (I_R^{88})^{-1/2}$. The red circles in Fig. 2(b) show the isotope shift $\Delta \nu = \Delta \nu_0 + \kappa^{88}(\nu^{88})I_R^{88} - \kappa^{87}(\nu^{87})I_R^{87}$ that linearly depends on the clock intensity $I_R^{88}$, as the clock-light shift $\delta\nu_R^{88} = \kappa^{88}(\nu^{88})I_R^{88}$ is about $10^5$ times larger than $\delta\nu_R^{87} = \kappa^{87}(\nu^{87})I_R^{87} \approx -40$ µHz; therefore, we neglect $\delta\nu_R^{87}$ in the following discussion. To compensate the clock-light shift $\delta\nu_R^{88}$, while interrogating $^{87}$Sr, we simultaneously apply the clock laser with the intensity $I_R^{88}$ and the frequency $\nu^{88}$, which is off-resonant by $f_0 = 62.188$ MHz from $\nu^{87}$. This successfully compensates the clock-light shift as $\Delta \nu = \Delta \nu_0 + \kappa^{88}(\nu^{88})I_R^{88} - \kappa^{87}(\nu^{88})I_R^{88}$ shown by black circles. The blue solid line shows a linear fit to the black circles, indicating a fractional differential polarizability $\Delta\kappa/|\kappa| = [\kappa^{88}(\nu^{88}) - \kappa^{87}(\nu^{88})]/|\kappa^{87}| = 1(2)\times10^{-3}$ consistent with zero. Note that the absolute values[5, 18] cited for $\beta^A$ and $\kappa^A$ negligibly affect the isotope shift because of the excellent cancellation of perturbations, as described above.

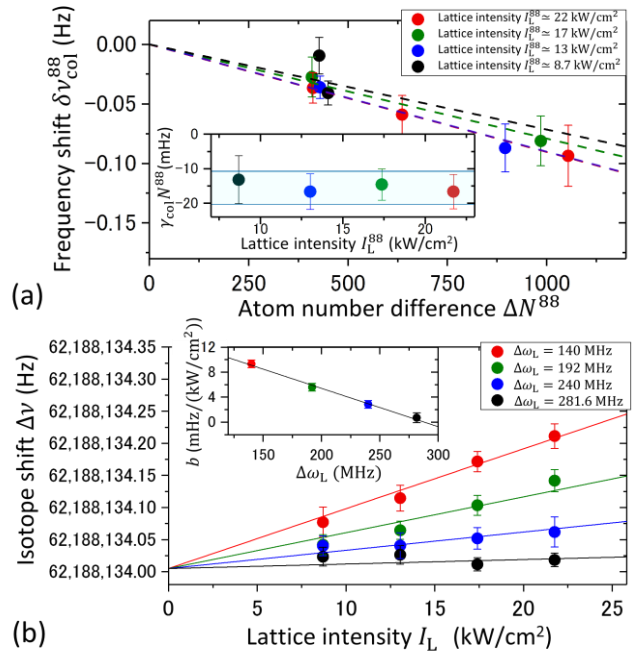

**Fig. 3.** (a) Frequency shift $\delta\nu_{col}^{88}(\Delta N^{88}, I_L^{88})$ of the $^{88}$Sr clock measured for different atom numbers $\Delta N^{88} = N_2^{88} - N_1^{88}$ and lattice intensities $I_L^{88}$. The dashed lines show a linear fit to the data, where the blue and red lines are nearly

overlapped. The inset shows the collisional shift for a typical number of atoms $N^{88} \approx 180(50)$. (b) Isotope shift measured for different lattice intensities $I_L$ and for lattice laser frequencies $\Delta\omega_L = \omega_L^{88} - \omega_L^{87}$ with $\omega_L^{87} = 368\ 554\ 490(3)$ MHz. The solid lines show a linear fit to the isotope shift with a common $y$-intercept. The derived slopes $b(\Delta\omega_L) = a[\Delta\omega_L - (\omega_{E1}^{88} - \omega_{E1}^{87})]$ are summarized in the inset in corresponding colors. The error bars represent the $1\sigma$ statistical uncertainty.

The collisional shift for $^{88}$Sr is much larger than that for $^{87}$Sr, because of the presence of the $s$-wave collisions that are suppressed for spin-polarized $^{87}$Sr owing to Pauli suppression[23]. We investigate the collisional shift by the interleaved operation of the clock alternating between low $N_1^{88}(\approx 200 - 300)$ and high $N_2^{88}(\approx 600 - 1300)$ atom numbers at different lattice intensities $I_L^{88}$ that also result in changes to the atomic density. Figure 3(a) summarizes the frequency shift $\delta\nu_{\text{col}}^{88}(\Delta N^{88}, I_L^{88}) = \nu^{88}(N_2^{88}, I_L^{88}) - \nu^{88}(N_1^{88}, I_L^{88})$ with $\Delta N^{88} = N_2^{88} - N_1^{88}(\approx 400 - 1{,}000)$. Assuming a linear dependence, $\delta\nu_{\text{col}}^{88}(\Delta N^{88}, I_L^{88}) = \gamma_{\text{col}}(I_L^{88})\Delta N^{88}$, we fit the data points with corresponding colors to derive $\gamma_{\text{col}}(I_L^{88})$. The inset shows the collisional shift $\gamma_{\text{col}}(I_L^{88})N^{88} \approx -15.5(4.8)$ mHz for a typical number of atoms, $N^{88} \approx 180(50)$. Since no clear lattice intensity dependence is observed within the measurement uncertainty, we evaluate the collisional shift using the weighted average and its uncertainty using the root-sum-squared (RSS) of the fitting error and the uncertainty (30%) of the number of atoms.

The lattice-light shift[24] is described as $\delta\nu_L^A(\omega) = \Delta\alpha_{E1}^A(\omega)I_L^A(T^A) + \delta\nu_h^A(I_L^A, T^A)$, where $\Delta\alpha_{E1}^A(\omega) = \left.\frac{\partial\Delta\alpha_{E1}^A(\omega)}{\partial\omega}\right|_{\omega=\omega_{E1}^A}(\omega - \omega_{E1}^A)$ is the difference in the

electric-dipole (E1) polarizability in the clock transition near the E1-magic frequency $\omega_{E1}^A$, $I_L^A$ is the effective lattice intensity dependent on the atomic temperature $T^A$, which determines the spatial distribution of atoms in the lattice, and $\delta\nu_h^A$ is the sum of the higher order lattice-light shifts including the magnetic-dipole, electric-quadrupole and hyper polarizability effects. Since we operate the clock with the same lattice laser intensity and similar atom temperatures $T^{87} \approx T^{88}$ for both isotopes, we assume equal effective intensities, $I_L^{87} \approx I_L^{88} \equiv I_L$. In addition, because the isotope shift and the hyperfine splitting of about 1 GHz are much smaller than the detuning ($> 60$ THz) of the lattice laser from the atomic resonance, we assume[24] a frequency dependence of the E1-polarizability $\left.\frac{\partial \Delta\alpha_{E1}^A(\omega)}{\partial \omega}\right|_{\omega=\omega_{E1}^A} \equiv a$ and that higher order light shift contributions $\delta\nu_h^A(I_L^A, T^A)$ are independent of isotopes $A = 87$ and $88$. Consequently, the cancellation of higher order lattice-light shift between the isotopes allows the differential lattice-light shift to vary linearly with the lattice laser intensity $I_L$, i.e., $\Delta\nu = \Delta\nu_0 + \delta\nu_L^{88}(\omega_L^{88}) - \delta\nu_L^{87}(\omega_L^{87}) = \Delta\nu_0 + a[\Delta\omega_L - (\omega_{E1}^{88} - \omega_{E1}^{87})]I_L$. Note that the frequency shift then depends only on the frequency difference $\Delta\omega_L = \omega_L^{88} - \omega_L^{87}$ of the lattice frequency applied for each isotope. The circles in Fig. 3(b) show the isotope shift $\Delta\nu$ as a function of the lattice intensity $I_L$ for $\Delta\omega_L = 140, 192, 240$, and $281.6$ MHz, where we set $\omega_L^{87} = 368\ 554\ 490\ (3)$ MHz, as reported in Ref. 22. The inset shows the corresponding linear slopes $b(\Delta\omega_L) = a[\Delta\omega_L - (\omega_{E1}^{88} - \omega_{E1}^{87})]$ as a function of $\Delta\omega_L$, which determines the isotope shift $\omega_{E1}^{88} - \omega_{E1}^{87} = 288\ (7)$ MHz of the E1 magic frequency. Here, the uncertainty is given by the RSS of the fitting error and the collisional shift uncertainty.

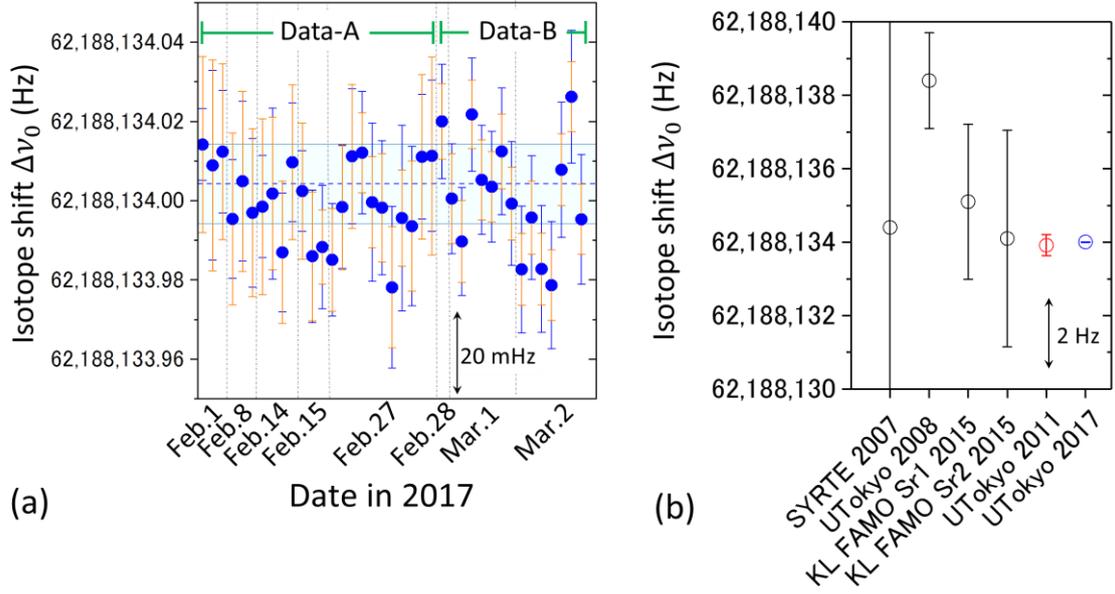

**Fig. 4** (a) Isotope shift measurements carried out over one month. Data-A are used to investigate the systematic uncertainties. By averaging 15 measurements after 28 February 2017 (data-B), the isotope shift is measured to be $\Delta\nu_0 = 62\ 188\ 134.004(10)$ Hz, as given by the blue dashed line with the shaded region indicating uncertainty. The blue and orange error bars represent the $1\sigma$ statistical uncertainty and the systematic uncertainty, respectively. (b) Summary of isotope shift measurements. The present result (blue circle) is consistent with our latest measurement in 2011 (red circle, unpublished) performed with the experimental setup reported in Ref. 25. The error bars represent the total uncertainty.

Figure 4(a) summarizes the isotope shift measurements (blue circles) obtained over one month, February–March 2017. The data set before 27th February (data-A) is used to investigate the systematic effects; therefore, each point has different systematic uncertainty (orange bars). The measurements after the 28th (data-B) are performed at fixed parameters except for the number of atoms, which reduces the systematic uncertainties. Using the latter data set (data-B), the weighted average according to the statistical uncertainty of each data point (blue bars) is used to determine the isotope shift to be $\Delta \nu_0 = 62\ 188\ 134.004(10)$ Hz. The total uncertainty of 10 mHz comprises the statistical uncertainty of 4 mHz, calculated from the standard error of the mean, and the systematic uncertainty of 9 mHz, calculated from the weighted mean of the systematic uncertainty for each data point. To verify the reproducibility, we evaluate the isotope shift using the data-A, which gives a consistent result of $\Delta \nu_0 = 62\ 188\ 134.004(18)$ Hz. Table I summarizes the systematic corrections and uncertainties for $^{88}$Sr, $^{87}$Sr, and the isotope shift. Thanks to the cancellation of perturbations, the systematic uncertainties for the isotope shift are smaller than those of individual transition frequencies except for collisional shift. Figure 4(b) summarizes the values of the isotope shift reported so far[13-16], where we subtract the recommended frequency $\nu(^{87}$Sr$)$ in 2015[12] from the absolute frequencies reported in Refs. 15 and 16. Regrettably, our value of 62 188 138. 4(1.3) Hz measured in 2008[13] disagrees with the current result; we later found that the previous measurements were affected by the lattice-light shift resulting from the amplified spontaneous emission (ASE)[16] of a semiconductor amplifier used for one of the lattice lasers. A red circle shows the isotope shift $\Delta \nu_0 = 62\ 188\ 133.9(0.3)$ Hz that we reevaluated in 2011 using the experimental setups reported in Ref. 25 with the careful removal of ASE, and it is consistent with the present measurement that is nearly free of the common ASE effect.

Table I. Systematic corrections and uncertainties for $^{88}$Sr, $^{87}$Sr, and the isotope shift. The blackbody radiation shift is estimated for $T = 297.1$ K of the surrounding vacuum chamber, using parameters reported in Refs. 4 and 27. An acousto-optic modulator (AOM) was used as a frequency shifter (FS1).

| Contributors | $\nu(^{88}Sr)$ | | $\nu(^{87}Sr)$ | | $\nu(^{88}Sr) - \nu(^{87}Sr)$ | |
| :---: | :---: | :---: | :---: | :---: | :---: | :---: |
| | Corr. (mHz) | Unc. (mHz) | Corr. (mHz) | Unc. (mHz) | Corr. (mHz) | Unc. (mHz) |
| Quadratic Zeeman shift | 17096.0 | 205.7 | 17096.8 | 205.6 | -0.8 | 4.0 |
| Lattice-light shift | -2.9 | 6.7 | 2.3 | 3.2 | -5.1 | 5.8 |
| Clock-light shift | 1614.6 | 19.5 | 1616.2 | 19.9 | -1.6 | 3.6 |
| Collisional shift | 15.5 | 4.8 | 0.0 | 0.2 | 15.4 | 4.8 |
| Servo error | 0.0 | 3.9 | -1.4 | 3.9 | 1.4 | 1.3 |
| Common effect | | | | | | |
| Blackbody radiation shift | 2188.7 | 138.7 | 2188.7 | 138.7 | 0 | < 0.1 |
| AOM chirp | 0 | 0.1 | 0 | 0.1 | 0 | < 0.1 |
| Systematic total | 20911.9 | 249.0 | 20902.6 | 248.9 | 9.3 | 9.4 |

In summary, we demonstrated the rejection of perturbations in the isotope shift measurement by alternately interrogating the $^{88}$Sr and $^{87}$Sr transitions in the presence of the same optical and magnetic perturbations within the same experimental setup. Because of the reduction of the uncertainties originating from clock-light shift, quadratic Zeeman shift, and blackbody radiation shift, we determined the isotope shift to be $\Delta\nu_0 =$ 62 188 134.004(10) Hz, which allows us to relate the $^{88}$Sr clock transition frequency to that of $^{87}$Sr with a fractional uncertainty of $2\times10^{-17}$. The same technique can be applied to other optical clocks operated with multiple isotopes, such as Yb, Hg, and Cd optical lattice clocks. The efficient cancellation of the perturbations allows the precise determination of the isotope shift that may serve as a probe of the atomic Higgs force[26].

**Acknowledgments**

This work was supported by JST ERATO Grant Number JPMJER1002 10102832, by JSPS Grant-in-Aid for Specially Promoted Research Grant Number JP16H06284, and by the Photon Frontier Network Program of the Ministry of Education, Culture, Sports, Science and Technology (MEXT), Japan. We thank N. Ohmae, M. Takamoto, A. Yamaguchi, and N. Nemitz for careful reading of the manuscript.